# Biological effects and equivalent doses in radiotherapy: a software solution

Cyril Voyant,[1,2*] Daniel Julian,[3] Rudy Roustit,[4] Katia Biffi,[2] and Céline Lantieri[2]

1-University of Corsica, CNRS UMR SPE 6134, (Campus Grimaldi, 20250 Corte), France

2-Hospital of Castelluccio, Radiotherapy Unit, BP 85, 20177 Ajaccio, France

3-Joseph Fourier University, 38000 Grenoble, France

4- Centre de la république, Radiotherapy Unit, 63000 Clermont-Ferrand

*Corresponding author: Cyril Voyant

Email: voyant@univ-corse.fr; Tel.: +33 495293666; Fax: +33 495293797

## Abstract

The limits of TDF (time, dose, and fractionation) and linear quadratic models have been known for a long time. Medical physicists and physicians are required to provide fast and reliable interpretations regarding the delivered doses or any future prescriptions relating to treatment changes. We therefore propose a calculation interface under the GNU license to be used for equivalent doses, biological doses, and normal tumor complication probability (Lyman model). The methodology used draws from several sources: the linear-quadratic-linear model of Astrahan, the repopulation effects of Dale, and the prediction of multi-fractionated treatments of Thames. The results are obtained from an algorithm that minimizes an ad-hoc cost function, and then compared to the equivalent dose computed using standard calculators in seven French radiotherapy centers.

# Nomenclature

α et β   fitting parameters of the linear quadratic model of cell survival (Gy² et Gy)

$\alpha_{unsc}$   adjustment parameter of the occurrence model of cancer radio-induced (Gy$^{-1}$)

$\theta(x)$   Heaviside function

$\frac{\gamma}{a}$   parameter of the LQL model

μ   parameter adjustment necessary to take into account the poly-fractionation in the model LQ (hours$^{-1}$)

BED   biological equivalent dose (Gy)

D   physical dose (Gy)

dt   dose per fraction from which the curve of cell survival becomes linear (Gy)

$D_{prol}$   proliferation dose (Gy/day)

$D_1$ et $D_2$ equivalent doses for the treatment 1 and 2 (Gy)

$EQD_2$   equivalent dose for a 2 Gy/Fraction treatment (Gy)

EUD   Equivalent uniform Dose (Gy)

$EUD^{2Gy}$ EUD for an equivalent dose related to a reference of 2 Gy per fraction

$f$   cost function to minimize by the algorithm

$ja$   number of day-offs

$H_m$   LQ model correction taking account the poly-fractionation

$K_{incidence}$ occurrences probability of radio induced cancer (%)

$m$   fraction number and slope factor of the NTCP model

$n$   nombre de fraction

NTCP   complications rate of post radiation (%)

$P_{unsc}$   parameter related to the occurrence of radiation-induced cancers (Gy$^{-1}$)

ΔT   duration between two irradiations (heures)

T   overall time (day)

TD50 dose at which there is a 50% complication (Gy)

$T_k$   time at which repopulation begins after start of treatment (day)

$T_{pot}$   potential doubling time (day)

$T_{stop}$   days off during the treatment

$u$   boundary used in the NCTP calculus (Gy)

# I. Introduction: Problems of the biologically equivalent dose

It has long been known that radiation biology plays an important role and is necessary for radiotherapy treatments. The radiation effects on normal and malignant tissues after exposure range from a femtosecond to months and years thereafter [1,2]. Therefore, to optimize treatment, it is crucial to explain and understand these mechanisms [3-5]. Providing a conceptual basis for radiotherapy and identifying the mechanisms and processes that underlie the tumor and normal tissue responses to irradiation can help to explain the observed phenomena [6]. Examples include understanding hypoxia, reoxygenation, tumor cell repopulation, or the mechanisms of repair of DNA damage [3,7,8]. The different biological effects of radiation should be divided into several phases: the physical phase (interaction between charged particles and tissue atoms), chemical phase (the period during which the damaged atoms and molecules react with other cellular components in rapid chemical reactions), and biological phase (impact of the generated lesions on the biological tissue [4]). The following section describes the models most often used in radiotherapy. These are simplistic models that actual treatments are based, and that are validated and approved [9-12].

## 1. Reference models

Numerous models exist to evaluate the biological equivalent dose, but the two most common are the nominal standard dose (NSD [13]) and linear quadratic (LQ [9]) models. The NSD uses the power law described in equation 1 below ($D_{tol}$ is the tolerance dose of the tissue, *NSD* is a constant, *n* and *t* $\in \mathbb{R}^+$, *N* the number of fractions, and *T* the overall treatment time). However, this model has been often criticized [14]. In short, some researchers consider and have even shown that the NSD formula is not a valid description for all tumors and normal tissues; instead, they maintain that the model incorrectly describes the effects of fraction number and treatment duration.

$$D_{tol} = NSD \cdot N^n \cdot T^t \qquad \text{Eq 1}$$

The LQ model is most frequently used in the radiotherapy units. It allows the equivalent dose to be easily evaluated for different fractionations. This concept involves the $\frac{\alpha}{\beta}$ ratio, as shown in equation 2 below (*D* is the total dose for a fraction size of *d* gray).

$$EQD_2 = D \cdot \frac{d + \frac{\alpha}{\beta}}{2 + \frac{\alpha}{\beta}} \qquad \text{Eq 2}$$

$EQD_2$ is the dose obtained using a 2Gy fraction dose, which is biologically equivalent to the total dose *D* given with a fraction dose of *d* gray. The values of $EQD_2$ may be added in separate parts in the treatment plan. This formula may be adapted to fraction doses other than 2Gy.

## 2. Limitations of the LQ model

The LQ model is frequently used for modeling the effects of radiotherapy at low and medium doses per fraction for which it appears to fit clinical data reasonably well. The main disadvantage of the LQ approach is that the overall time factor is not taken into account, because in radiotherapy it is regarded to be more complex than previously supposed [3]. It is indeed very difficult to include this parameter in the LQ equation. However, a technique may be used to integrate a penalty term in Equation 2. Thus, for $T_{stop}$ days off treatment, the dose recovered would be $T_{stop} \cdot D_{prol}$, where $D_{prol}$ is the proliferation factor (in Gy/day; for example, 0.22 for laryngeal edema or 0.15 for rectosigmoid complications). This methodology is essentially validated for discontinuation during treatment. As a general rule, the main limitations of using the LQ model are linked to repopulation (LQ doesn't take into account the dose protraction), bi-fractionated treatments and high-dose fractions (continuously bending survival curve versus linear behavior observed at least in some cell lines). Other more sophisticated models, however, take into account these weaknesses. We will later see that the LQ model requires further theoretical investigation, especially in terms of the biologically effective dose (BED).

Given the difficulty of computing the BED, we conducted a study in seven radiotherapy centers in France: CHD Castelluccio (Ajaccio; two classical calculators used), Centre de Cancérologie du Grand

Montpellier (Montpellier), CRLCC Paul Lamarque (Montpellier), Clinique Saint-Pierre (Perpignan), Centre de la République (Clermont Ferrand), CHU of Grenoble, and CHU of Nîmes. A questionnaire was sent to medical physicists working at these centers, with the aim of comparing the results of equivalence (for standard radiotherapy planning). Table 1 presents the results of this survey, which indicate that not all of the operators obtained the same results. The 95 % confidence interval was often very large. Moreover, the relative standard deviation (also known as the dispersion coefficient) was frequently greater than 5% (13 times out of 24). This dispersion was larger in the case of target volumes; the maximal volume (close to 40%) was related to high doses per fraction with a gap between two radiotherapy cycles. In Table 1, it is evident that all of the users did not estimate dose equivalence in the same manner. Only for a dose per fraction approaching 2Gy and standard overall time were the results equivalent. Note that in the multi-fractionated treatments, only 50% of the centers were able to give an equivalent dose, as this kind of treatment was not computable.

| Treatements | | Organs at risk | Target volumes |
|---|---|---|---|
| | | Spinal cord | Prostate (metastasis) |
| 10x3Gy | Median | 37.50 | 36.65 |
| | average ± 95% CI | 37.8 ± 1.2 | 36.6 ± 1.9 |
| | stand dev | 1.71 / 4.5% | 2.87 / **7.8%** |
| | | Spinal cord | Breast (metastasis) |
| 10x3Gy | median | 37.50 | 35.57 |
| | average ± 95% CI | 37.8 ± 1.2 | 35.97 ± 1.4 |
| | stand dev | 1.71 / 4.5% | 2.06 / **5.7%** |
| | | Spinal cord | Prostate (metastasis) |
| 1x8Gy | median | 20.00 | 14.90 |
| | average ± 95% CI | 19.1 ± 2.3 | 15.9 ± 3.2 |
| | stand dev | 3.38 / **17.7%** | 4.72 / **29.6%** |
| | | Brain | Breast (metastasis) |
| 10x3Gy | median | 37.50 | 35.57 |
| | average ± 95% CI | 37.5 ± 0.9 | 35.9 ± 1.4 |
| | stand dev | 1.26 / 3.4% | 2.03 / **5.6%** |
| 1x8Gy | | Spinal cord | Prostate (metastasis) |
| (1 month gap time) | median | 33.30 | 21.50 |
| 1x8Gy | average ± 95% CI | 33.6 ± 4.0 | 24.04 ± 6.8 |
| | stand dev | 5.78 / **17.2%** | 9.76 / **40.6%** |
| | | Pericardium | Lung (metastasis) |
| 5x4Gy | median | 30.90 | 27.07 |
| | average ± 95% CI | 33.7 ± 5.9 | 28.9 ± 3.6 |
| | stand dev | 8.47 / **25.1%** | 5.14 / **17.8%** |

| | | Oral mucosa | Oropharynx |
|---|---|---|---|
| 20x2Gy (1 week gap time) 10x2Gy | median | 57.95 | 57.95 |
| | average ± 95% CI | 58.1 ± 0.5 | 57.6 ± 1.9 |
| | stand dev | 0.66 / 1.14% | 2.76 / 4.8% |
| | | Oral mucosa | Oropharynx |
| 22x1.8Gy (bi-fractionated) | median | 41.95 | 41.70 |
| | average ± 95% CI | 41.0 ± 1.6 | 41.7 ± 2.2 |
| | stand dev | 2.32 / **5.6%** | 3.12 / **7.5%** |
| | | Rectum | Prostate |
| 25x1.8Gy then 15x2Gy | median | 72.58 | 72.75 |
| | average ± 95% CI | 72.6 ± 0.4 | 72.6 ± 0.7 |
| | stand dev | 0.54 / 0.7% | 1.00 / 1.37% |
| | | Lung | Breast |
| 20x2.5Gy (4 fraction/week) | median | 55.58 | 53.50 |
| | average ± 95% CI | 55.0 ± 1.1 | 53.5 ± 1.2 |
| | stand dev | 1.59 / 2.9% | 1.71 / 3.2% |
| | | Optic chiasma | Glioblastoma |
| 4x4.5Gy (2 week gap time) 4x4Gy | median | 50.25 | 42.15 |
| | average ± 95% CI | 49.9 ± 2.7 | 43.1 ± 2.8 |
| | stand dev | 3.96 / **7.9%** | 4.07 / **9.4%** |
| | | Skin (early) | Breast |
| 28x1.8Gy (1 week gap time) | median | 46.65 | 46.36 |
| | average ± 95% CI | 46.7 ± 0.6 | 46.5 ± 0.8 |
| | stand dev | 0.90 / 1.9% | 1.13 / 2.4% |

Table 1: Methodologies for computing the equivalent dose used in eight clinical calculators from seven radiotherapy centers. The median dose, average dose, and standard deviation are given in Gy. For the standard deviation, absolute and relative modes ($\sigma$/average) were used. Bold font is used to represent values >5%

The numbers of centers included in this study was low, and there was no consensus among the centers in terms of their methods for computing the doses. If we look more closely, the biological equivalent dose was the only process that was calculated using non-official software. All of the other steps in treatment planning followed an official protocol. It is thus legitimate to ask why centers use in-vivo dosimeters or try to achieve a global error of 2% throughout treatment, if the prospective calculation of the equivalent dose (and prescription) is greater than 20%. In order to address the question of responsibility, the following section of this article is targeted at medical physicists, knowing that the prescriber is the physician. In case of equivalent computations, the optimal operation would be for the technical work to be performed by the physicist and validations by the physician

(while taking into account the clinical scenario). This methodology allows for a double checking of the calculation results.

The next section describes the theoretical methodology that we propose to compute the BED.

## II. Materials and methods: the developed models

The BED (introduced by Fowler [9]) is a mathematical concept used to illustrate the biological effects observed after irradiation. In addition to being easily computable (BED = physical dose x relative efficiency), this notion is interesting because two irradiations with the same BED generate the same radiobiological effects. For this reason, it is easy to compare treatments with different doses, fractionations, and overall times. The following section introduces the BED-based models that we advocate as well as the rules and guidelines for using the LQL_equiv software.

### 1. Target volume models

Let us examine two different treatment cases separately. The first involves treatments with a high-dose fraction (one treatment per day, the fraction size $d$ is greater than the $d_t$ limit; [15]), which requires a linear quadratic linear (LQL) model. The second case relates to other treatments ($d < d_t$), where the LQ model is applicable to daily multi-fractionation [16].

#### a. The $d > d_t$ case

When the dose per fraction ($d$) is greater than the LQL threshold ($d_t \sim 2.\alpha/\beta$), the BED is computed using Equation 3 below (one fraction permitted per day). This template regroups Astrahan's high-dose model [17] and Dale's repopulation model [18] ($n$ is the number of fraction, $\theta(x)$ the Heaviside function, $\frac{\gamma}{a}$ the parameter of the LQL model and $T_{pot}$ the potential doubling time in day).

$$BED = n.\left(d_t.\left(1 + \frac{d_t}{\alpha/\beta}\right) + \frac{\gamma}{a}.(d - d_t)\right) - \theta(T - T_k).\frac{\ln(2)}{\alpha.T_{pot}}.(T - T_k) \qquad \text{Eq 3}$$

The second term used in this equation is useful only when the overall time $T$ is greater than the $T_k$ value (kick-off time). If this threshold is not achieved, the tumor is considered to be non-proliferative (early hypoxia).

### b. The $d \leq d_t$ case

When the fraction dose is low, it is possible to use the standard BED equations while considering one or more fractions per day (Equation 4). This methodology follows the model of Thames [19], who introduces the repair factor $H_m$ related to the amount of unrepaired damage (Equation 5). If the inter-fraction interval is reduced below the full repair interval (between 6 hours and 1 day), the overall damage from the whole treatment is increased because the repair of damage due to one radiation dose may not be complete before the next fraction is given ($H_m$ is LQ model correction taking account the poly-fractionation, $m$ the number of fraction per day, $\phi$ the incomplete repair and $\mu$ the parameter adjustment necessary to take into account the poly-fractionation in the model LQ in hours$^{-1}$).

$$BED = n.d.\left(1 + (1+H_m).\frac{d}{\alpha/\beta}\right) - \theta(T-T_k).\frac{\ln(2)}{\alpha.T_{pot}}.(T-T_k) \qquad \text{Eq 4}$$

$$H_m = \left(\frac{2}{m}\right).\left(\frac{\phi}{1-\phi}\right).\left(m - \frac{1-\phi^m}{1-\phi}\right) \text{ and } \phi = e^{(-\mu.\Delta T)} \qquad \text{Eq 5}$$

Note that in the case of mono-fractionation, the $H_m$ factor is null. These equations only relate to the target volume calculation. For the organs at risk, the kick-off time is not relevant, meaning that it is necessary to use a repopulation-specific approach.

### 2. Models for organs at risk

As in the precedent section on target volumes, this section similarly separates high and low doses per fraction. The BED formulae are almost equivalent to the target volume model; only the terms relating to the lack of dose by proliferation are modified.

### a. The d > $d_t$ case

To understand this methodology, it is necessary to consult Van Dyk's law [20]. The kick-off time is no longer considered, with the recovered dose ($D_{rec} = \frac{\ln(2)}{\alpha.T_{pot}}$ in Gy/day) instead being added. The global model is described in Equation 6 below.

$$BED = n.\left(d_t.\left(1 + \frac{d_t}{\alpha/\beta}\right) + \frac{\gamma}{a}.(d - d_t)\right) - D_{rec}.T \qquad \text{Eq 6}$$

### b. The d ≤ $d_t$ case

In the case of low doses per fraction, the methodology is similar to the target volume model: the $H_m$ parameter (Equation 5) is nonetheless required, which allows us to take into account more than one fraction per day. As seen in the Equation 7 below, the recovered dose is used as in the previous case.

$$BED = n.d.\left(1 + (1 + H_m).\frac{d}{\alpha/\beta}\right) - D_{rec}.T \qquad \text{Eq 7}$$

### 3. Computational methods for the equivalent dose

The standard models used for the equivalent dose as based on the LQ approach are easily exploitable. The main formulation of the model (Equation 2) can be obtained by considering the general formula described in the Equation 8 as follows.

$$D_1 = D_2.\frac{(\alpha/\beta + .d_2)}{(\alpha/\beta + .d_1)} \qquad \text{Eq 8}$$

This equation may be validated using BED methodology. Considering the BED of two treatments to be equal, it appears that a simple relation links the two overall doses, $D_1 (= n_1.d_1)$ and $D_2 (= n_2.d_2)$. The detail of this procedure is shown in the Equation 9 below.

$$BED_1 = n_1.d_1.\left(1+\frac{d_1}{\alpha/\beta}\right) = BED_2 = n_2.d_2.\left(1+\frac{d_2}{\alpha/\beta}\right) \qquad \text{Eq 9}$$

In the case of more sophisticated BED formulations, it is not easy to determine a simple formula linking the $D_1$ and $D_2$ doses, as recovery and repopulation significantly complicate the computational principle. Most of the existing software that uses the overall time correction does not calculate the equivalent dose; instead, it only provides the BED for the chosen treatments. In clinical use, it is more valuable for the physician or physicist to work with the equivalent dose in standard fractionation. In this context, the methodology used in the LQL_Equiv software is based on an innovative algorithm, which allows a cost function extremum to be determined based on BED modeling. To explain this methodology, it is necessary to consider two irradiations (Indices 1 and 2), which are defined by a fraction number (*n*), dose per fraction (*d*), and days of discontinuation (*ja*). The corresponding BED is noted as BED1 ($n_1,d_1,ja_1$) and BED2 ($n_2,d_2,ja_2$), while the cost function *f* is defined in Equation 10 as follows.

$$f(n_1,d_1,ja_1,n_2,d_2,ja_2) = |BED_1(n_1,d_1,ja_1) - BED_2(n_2,d_2,ja_2)| \qquad \text{Eq 10}$$

In clinical use, it is desirable to compare a radiotherapy trial with one that is performed in a conventional manner (generally with 2 Gy per fraction without interruption). This concept of a reference dose simplifies the issue, as it is thus possible to dispense with the days off treatment and multi-fractionation per day in relation to the reference treatment. The following example concerns a tumor case with a dose per fraction less than $d_t$ (second part of the target volume model), while the cost function, *f*, is given in Equation 11. Concerning the three other cases examined in previous sections, a similar relationship is found.

$$f(n_{ref},d_{ref},n,d,ja) = \Bigg| n_{ref}.d_{ref}.\left(1+\frac{d_{ref}}{\alpha/\beta}\right) - \theta(T_{ref}-T_k).\frac{\ln(2)}{\alpha.T_{pot}}.(T_{ref}-T_k) - n.d.\bigg(1+$$
$$(1+H_m).\frac{d}{\alpha/\beta}\bigg) - \theta(T-T_k).\frac{\ln(2)}{\alpha.T_{pot}}.(T-T_k)\Bigg| \quad \text{Eq 11}$$

The global treatment duration can be seen to be directly associated with the fraction number and days off during radiotherapy. Following Equation 11, the 2Gy-per-fraction equivalent dose (EQD$_2$) for standard treatment with the characteristics $n, d, ja$ is given by the algorithm shown in Equation 12.

$$\begin{cases} \text{argmin}_{n_{ref} \in \mathbb{R}^+} f(n_{ref}, 2, n, d, ja) = n_0 \\ \qquad EQD_2 = 2.n_0 \end{cases} \qquad \text{Eq 12}$$

All of the results obtained in this section were implemented using a Matlab® standalone application known as LQL-equiv. The characteristics of this software, its limitations, and guidelines for its use are discussed in the following section.

## III. Results: LQL_Equiv software

The LQL_Equiv software was developed in collaboration by the CHD Castelluccio radiotherapy unit in Ajaccio and the University of Corsica. It is a free software released under the GNU license. The source codes, executable file, help files, and license terms are available at http://cyril-voyant.univ-corse.fr/LQL-Equiv_a34.html. Before installing this software, it is advisable to refer to the installation guide and to download and execute Matlab Component Runtime (MCR 32 bits, version 7.15 or later). This latter step is necessary since the application was programmed using the GUI Matlab® software (32 bits, v. 7.12) and deployed with the Matlab Compiler® (v. 4.12), which use MCR (a standalone set of shared libraries enabling the execution of Matlab® applications on a computer without an installed version of Matlab®). Users of the LQL_Equiv software are advised to provide us with comments on the software, its libraries (biological parameters for each organ or tumor type), or any bugs so as to allow us to develop the software. Note that the application requires Microsoft Windows® (the resolution and colors are for Vista or later versions).

### 1. Software

The graphical interface of the LQL_Equiv software software is presented in the Figure 1, divided into five sections: demographical zone, tissue choice (organs at risk and target volumes), reference zone (characteristics for computing the equivalent dose), treatment planning zone (three juxtaposed and independent treatments), and finally, the equivalent dose under the reference conditions. Prior to using the software, it is important to understand that repopulation or a high dose per fraction can considerably alter the standard equivalent results. Therefore, it is recommended for each user to verify the results obtained and validate them during an initial test phase. The results must be consistent with routine procedures as well as the data in the literature. The detail of the instructions allowing to use the software is available in the annex part.

Figure 1: Graphical interface for the LQL_Equiv software

The ideal scenario would be to compare these results with other software and obtain a mean score for the two outputs or for the outputs that minimize the physical dose. We recommend using this software as a secondary BED calculator. It aims to provide assistance, but cannot be used as a substitute for routine calculations made by a professional. The creators of the LQL_Equiv software cannot be held responsible for any errors caused by the misuse of the results obtained.

## 2. Comparison with standard models

This section compares the results of the LQL_Equiv software with the available clinical models. However,, it is important to note that all of the parameters used for calculating the equivalence are available on the graphical interface. Using Matlab™ and the downloadable source codes, it is easy to modify or complete these parameters. It is also possible to contact the software authors to assist in developing the software. LQL_Equiv is in direct competition with TDF Plan developed by Eye Physics LLC, which proposes a multitude of parameters. However, the software is dedicated to the calculation of BED and is not really consistent with the reference equivalent dose. Moreover, we aimed to develop ergonomic software with minimum of adjustable parameters, which ultimately complicate the interpretation of the output. These two approaches are nevertheless complementary; for more information about the different models used, refer to the TDF Plan website (http://www.eyephysics.com/tdf/Index.htm). Table 2 presents a comparison between outputs of the standard calculation models described in section II (LQ without proliferation and $\alpha/\beta$ =10 for oral mucosa and 2 for others) and the LQL-equiv software. The difference between the two approaches is substantial. The overall time effect and unusual doses per fraction result in completely different outputs. The maximum difference is close to 25%; this value is linked to the cell repopulation of prostate cancer. In this case, the non-specific methods are certainly not usable.

| Treatements | | Organs at risk | Target volums |
|---|---|---|---|
| 10x3Gy | | Spinal cord | Prostate (metastasis) |
| | classical output (Gy) | 37.5 | 37.5 |
| | LQL-equiv output (Gy) | 37.5 | 36 |
| | difference (Gy / %) | -0 / -0% | -1.5 / -4% |
| 10x3Gy | | Spinal cord | Breast (metastasis) |
| | classical output (Gy) | 37.5 | 37.5 |
| | LQL-equiv output (Gy) | 37.5 | 38.2 |
| | difference (Gy / %) | -0 / -0% | 0.7 / 1.9% |
| 1x8Gy | | Spinal cord | Prostate (metastasis) |
| | classical output (Gy) | 20 | 20 |
| | LQL-equiv output (Gy) | 16 | 16.8 |
| | difference (Gy / %) | -4 / **-20%** | -3.2 / **-16%** |
| 10x3Gy | | Brain | Breast (metastasis) |
| | classical output (Gy) | 37.5 | 37.5 |
| | LQL-equiv output (Gy) | 43.5 | 38.2 |
| | difference (Gy / %) | 6 / **16%** | 0.7 / 1.9% |
| 1x8Gy (1 month gap | | Spinal cord | Prostate (metastasis) |
| | classical output (Gy)) | 40 | 40 |
| time) 1x8Gy | LQL-equiv output (Gy) | 32 | 33.3 |
| | difference (Gy / %) | -8 / -4.63% | -6.7 / **16.7%** |
| | | Pericardium | Lung (metastasis) |
| 5x4Gy | classical output (Gy) | 30 | 30 |
| | LQL-equiv output (Gy) | 37.5 | 23.3 |
| | difference (Gy / %) | 7.5 / **25%** | -6.7 / **-22.3%** |
| | | Oral mucosa | Oropharynx |
| 20x2Gy (1 week gap time) 10x2Gy | classical output (Gy) | 60 | 60 |
| | LQL-equiv output (Gy) | 54.4 | 53 |
| | difference (Gy / %) | -5.6 / **-9.3%** | -7 / **-11.7%** |
| | | Oral mucosa | Oropharynx |
| 22x1.8Gy (bi-fractionated) | classical output (Gy) | 38.9 | 38.9 |
| | LQL-equiv output (Gy) | 45 | 36 |
| | difference (Gy / %) | 6.1 / **15.7%** | -2.9 / **-7.4%** |
| | | Rectum | Prostate |
| 25x1.8Gy then 15x2Gy | classical output (Gy) | 72.7 | 72.7 |
| | LQL-equiv output (Gy) | 71 | 73 |
| | difference (Gy / %) | -1.7 / -2.3% | 0.3 / 0.4% |
| | | Lung | Breast |
| 20x2.5Gy (4 fraction/week) | classical output (Gy) | 56.2 | 56.2 |
| | LQL-equiv output (Gy) | 62.9 | 56.8 |
| | difference (Gy / %) | 6.7 / **11.9%** | 0.2 / 0.3% |
| | | Optic chiasma | Glioblastoma |
| 4x4.5Gy (2 week gap time) 4x4Gy | classical output (Gy) | 53.2 | 53.2 |
| | LQL-equiv output (Gy) | 42.8 | 47.4 |
| | difference (Gy / %) | -10.4 / **-19.5%** | -5.8 / **-10.9%** |
| | | Skin (early) | Breast |
| 28x1.8Gy (1 week gap time) | classical output (Gy) | 47.9 | 47.9 |
| | LQL-equiv output (Gy) | 47.6 | 42.3 |
| | difference (Gy / %) | -0.3 / 0.6% | -5.6 / **-11.7%** |

Table 2 : Comparison between the outputs of the LQL_Equiv and standard calculation models (LQ without proliferation and with $\alpha/\beta$ =10 for oral mucosa and 2 for others). Bold font is used to show differences >5%.

In addition, for the BED and equivalent calculations, the LQL_Equiv software allows two other parameters to be obtained, which may be useful in clinical practice: the normal tumor complication probability (NTCP; [22]) and the ratio of radiation-induced cancer after irradiation.

### 3. Others elements computed by the software

In the LQL_Equiv software, the bottom the interface is dedicated to the calculation of the NTCP and ratio of radiation-induced cancer. For the first parameter, the formula for its computation (only for normal tissues) is based on the Lyman model [22] as presented in the Equation 12 below (*TD50* is the dose at which there is a 50% complication in Gy, $u$ the boundary used in the NCTP calculus in Gy and m the slope factor). To use this formula, it is necessary to first compute the EUD (Niemerko [21]). However, in practice, this quantity is not feasible. It is instead possible to use the equivalent dose related to a reference dose of 2 Gy per fraction ($EQD_2 \approx EUD^{2Gy}$). However, the NTCP formalism is valid for 2±0.2 Gy/fraction. Moreover, the DVH must be used, in which case the equivalent dose refers to the average dose for the parallel organs or the maximal dose (D5%) for the serial organs.

$$\begin{cases} NTCP(n,d,ja) = \frac{1}{\sqrt{2\pi}} \int_{-\infty}^{u} e^{-t^2/2} \, dt \\ u = \frac{EUD^{2Gy}(n,d,ja) - TD50}{mTD50} \end{cases} \quad \text{Eq 13}$$

The second add-on in the software concerns the estimation of radiation-induced cancer. The theory used was developed by the United Nations Scientific Committee on the Effects of Atomic Radiation (UNSCEAR; http://www.unscear.org/unscear/fr/publications.html). The different meta-analyses of previous radiological incidents are used in this model. The ratio of radiation-induced cancer (in %) relating to normal tissue is provided in Equation 14 as follows ($\alpha_{unsc}$ is the adjustment parameter of the occurrence model of cancer radio-induced in Gy$^{-1}$, $P_{unsc}$ the UNSCEAR probability and $D^{2Gy}$ the equivalent dose for a 2 Gy/Fraction treatment in Gy).

$$K_{incidence} = P_{unsc}.D^{2Gy}.e^{-\alpha_{unsc}.D^{2Gy}} \quad \text{Eq 14}$$

Note that methods used to compute NTCP and $K_{incidence}$ are simplified; it is evident that interested readers must identify more specialized documents. These parameters are given as additional information.

## IV. Conclusion

In this article, we have exposed the compiling results of various published LQ model modifications, which have been modified to be better suited for specialized radiotherapy techniques such as hypo- or hyperfractionation. The LQ model was modified to take into account multi-fractionation, repopulation, high-dose fractions, and overall time. Moreover, we propose a software program (LQL_equiv), integrating all of these concepts regarding the main organs at risk or target volumes. Moreover, this free and easy-to-use software allows the NTCP to be calculated. Finally, this software permits the obtained results to be compared and validated against other "homemade" models, with the purpose of harmonizing practices in interested centers. However, it is essential to don't consider models as "general biological rules", parameters and outputs uncertainties can be very large; this phenomenon is related to the number of regression parameters (parsimony principle) and to the data snooping (e.g. failure to adjust existing statistical models when applying them to new datasets).

## V.     Acknowledgments

We would like to thank the following people for their contribution: Stéphane Muraro (Centre de Cancérologie du Grand Montpellier), Norbert Aillères and Sébastien Siméon (CRLCC Paul Lamarque; Montpellier), Vincent Plagnol (Clinique Saint-Pierre; Perpignan), Nicolas Docquière and Jean-Yves Giraud (CHU de Grenoble), and Bérengère Piron (CHU de Nîmes).

## VI.     Annex 1: Instructions for use

The number of modifiable parameters in the LQL_Equiv software is minimal, while the items required to complete a dose equivalent calculation are limited. Only the white boxes can be modified.

The upper left part of the interface is dedicated to patient demography (identity and pathology) and operator traceability. These parameters are not essential for initiating the calculation. Below this, the reference dose per fraction should be provided; by default, the dose is 2 Gy/fraction.

In the top-right of the interface, there are two dropdown menus related to the organs at risk and target volumes chosen by the operator to obtain the equivalent dose. Once these steps are completed, it is necessary to define the desired treatment plans. Only three plans are proposed, but the software is able to test more by integrating the overall results in a single treatment plan, such as the EQ1 (dose, days off, and number of fractions must be adjusted). The overall time must be verified or else there may be some imprecision in the final calculation. A null number of fractions or doses results in cancelling the calculation of the equivalent dose (the duration of the sequence does not contribute to the final output).

After selecting the treatment plan and clicking on the calculation button, the BED and equivalent doses are given. The page may be printed, or otherwise, there is the digital archiving solution based on the Windows™ print screen button.

When taking into account the days off, the weekend should not be considered; only discontinuations that occur during weekdays should be included. Beyond 20 days off from treatment, the algorithms are no longer valid. In the first approximation, the side of caution indicates that healthy tissues do not recover during the gap time. For the second cycle of radiotherapy that occurs a long time after the first, we must be vigilant with regard to the treated organs. In the case of the skin, for example, we may consider a duration of 2 to 5 years to be sufficient to negate any effects from the previous treatment (this is, however, invalid if the effects are already visible at the time of irradiation), while for the spinal cord, it must be considered, where possible, that there exists a dose memory, with the effects of gray radiation always being present. In this regard, the software takes into account that certain organs, such as spinal cord, have a low $D_{rec}$ in order to limit the consequences to the most critical organs. Moreover, it is necessary to consider all of the treatment phases if a dose equivalent is required for the second stage of a prostate disease. In this case, the first phase of the treatment must be considered, or otherwise, the kick-off time will not be correctly taken into account.

To avoid the dose overestimation, we recommend first calculating the dose equivalent for the organ, *i.e.*, the limiting factor, and then estimating the fractionation effect on the target volume.

For organs at risk, it is possible to use the nominal dose. Thus, in the case of the pelvis, for the first 45 Gy given in 25 fractions, the dose received by the rectum may be considered equal to 45 Gy. However, in order to optimize the methodology, it seems more reasonable to utilize a more detailed analysis. If the validation criterion is D30, the software should be completed according to the dose per fraction and number of fractions for the dose received by 30% of the rectum. It is also possible to use the average dose for parallel organs, maximal dose for serial organs, or simply the equivalent uniform dose (EUD) [21]. Another example illustrating the difference between the critical dose and nominal standard dose is based on spinal irradiation. If doses of 30 Gy in 10 fractions are delivered, this does not necessary mean that the spinal cord has received the entire dose. Dose volume histogram (DVH) analysis allows us to observe that the spinal cord received 32 Gy after the 10 fractions, which means that the equivalent dose is 10 fractions of 3.2 Gy, which significantly changes the results.

Furthermore, it should be added that in this software, as is often the case, the time between two irradiations in bi-fractionated radiotherapy must be greater than 6 hours.

## VIII. Conflicts of interest

The authors declare to have no conflicts of interest.